\documentclass[twocolumn,showpacs,preprintnumbers,amsmath,amssymb]{revtex4}

\usepackage{graphicx}
\usepackage{dcolumn}
\usepackage{bm}

\begin{document}

\preprint{APS/123-QED}

\title{Hyperfine interaction and magnetoresistance in organic semiconductors}

\author{Y. Sheng$^1$, D. T. Nguyen$^1$, G. Veeraraghavan$^2$, \"{O}. Mermer$^1$, M. Wohlgenannt$^1$} \email{markus-wohlgenannt@uiowa.edu}\author{U. Scherf$^3$}

\address{$^1$Department of Physics and Astronomy and Optical Science and Technology Center, University of Iowa, Iowa City, IA 52242-1479, USA}
\address{$^2$Department of Electrical and Computer Engineering and Optical Science and
Technology Center, University of Iowa, Iowa City, IA 52242-1595, USA}
\address{$^3$Bergische Universit\"at Wuppertal, Fachbereich Chemie, Makromolekulare Chemie, D-42097 Wuppertal, Germany}

\date{\today}

\begin{abstract}

We explore the possibility that hyperfine interaction causes the recently discovered organic magnetoresistance (OMAR) effect. Our study employs both experiment and theoretical modelling. An excitonic pair mechanism model based on hyperfine interaction, previously suggested by others to explain magnetic field effects in organics, is examined. Whereas this model can explain a few key aspects of the experimental data, we, however, uncover several fundamental contradictions as well. By varying the injection efficiency for minority carriers in the devices, we show experimentally that OMAR is only weakly dependent on the ratio between excitons formed and carriers injected, likely excluding any excitonic effect as the origin of OMAR.

\end{abstract}


\maketitle

\section{\label{sec:level1}Introduction}

Organic $\pi$-conjugated materials have been used to manufacture devices such as organic light-emitting diodes (OLEDs) \cite{Friend:1999,Forrest:2004}, photovoltaic cells \cite{Brabec:2001,Peumans:2003,Granstr:1998} and field-effect transistors \cite{Dimitrakopoulos:2002,Grundlach:1997,Shtein:2002}. Recently there has been growing interest in spin \cite{Wohlgenannt:2001,Dediu:2002,Xiong:2004,Hu:2006} and magnetic field effects (MFE) \cite{Francis:2004,Mermer:2005,Mermer:2005b,Preprint,Kalinowski:August2003,Kalinowski:2004,Davis:2004,Yoshida:2005,Salis:2004,Prigodin:2002,Raju:2003} in these materials. Frankevich and coworkers \cite{Frankevich:1992, Frankevich:1996} and Kalinowski and coworkers \cite{Kalinowski:June2003, Kalinowski:August2003} studied the effect of small magnetic fields on excitonic processes that occur in OLEDs, such as photoconductivity, electroluminescence (EL) and exciton dissociation at the electrodes. They have convincingly explained their findings using a model where the applied magnetic field reduces the effect of the hyperfine interaction between electron/hole spin and the hydrogen nuclei in the organic molecules. We will refer to this work as the excitonic pair mechanism model that will be treated in detail in section~\ref{subsec:pairmechanism model}.

We recently discovered \cite{Francis:2004} a large and intriguing magnetoresistive effect in OLEDs, which we dubbed organic magnetoresistance (OMAR). OMAR may find application in magnetic field sensors, e.g. in OLED interactive displays (patent pending, see demonstration video in Ref. \cite{Francis:2004}). In addition to its potential applications, OMAR poses a significant scientific puzzle since it is, to the best of our knowledge, the only known example of large room temperature magnetoresistance in non-magnetic materials with the exception of narrow-gap high-mobility materials \cite{Xu:1997}.

In the present work we examine whether the excitonic pair mechanism model \cite{Frankevich:1992,Kalinowski:June2003, Kalinowski:August2003} can explain OMAR. For this purpose we recast this model (see section~\ref{subsec:pairmechanism model}) into a form suitable for discussing the whole body of experimental MFE data, including both EL and transport measurements (OMAR). We find that whereas the excitonic pair mechanism model can explain several key aspects of experiments, it leads to several serious contradictions with experiment, especially in relation to the magnetotransport data. We trace the origin of these contradictions to the fact that this model in its present form has been applied to describe the spin-dynamics of neutral polaron pairs that do not significantly affect the current. Furthermore, we show experimentally that the magnitude of OMAR depends only weakly on the ratio, $\eta_{1}$, between excitons formed and charge carriers injected into the devices. In our opinion, this shows that any model based on excitonic processes fails to explain OMAR.

\section{Experimental}

\begin{figure}
 \includegraphics[width=0.75\columnwidth]{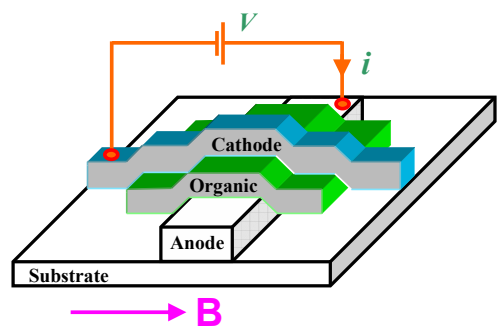}\\
 \caption{A schematic drawing of the device and the magnetoresistance experiment} \label{fig:DeviceSchematic}
\end{figure}

We first describe the sources for the various organic semiconductors we used in our study. The $\pi$-conjugated polymer polyfluorene (PFO) was purchased from American Dye Source, inc. The methyl-substituted ladder-type poly(p-phenylene) (MeLPPP) polymer was synthesized as described elsewhere \cite{Scherf:1991}. The $\pi$-conjugated small molecule Alq$_3$ was purchased from H. W. Sands corporation and was used as received.

The fabrication of the organic sandwich devices started with glass substrates coated with 40nm of indium-tin-oxide (ITO), purchased from Delta Technologies. The conducting polymer Poly (3,4-ethylenedioxythiophene)-poly (styrenesulfonate) (PEDOT), purchased from H. C. Starck was spin coated at 2000 rpm on top of the ITO to provide an efficient hole injecting electrode. All other manufacturing steps were carried out in a nitrogen glove box. The active polymer film was spin coated onto the substrate from a chloroform solution. The small molecular film layers were made by thermal evaporation. The cathode, either Ca (with an Al capping layer), Al, or Au, was then deposited by thermal (Ca) or electron beam evaporation (Al, Au) at a base pressure of $\approx 1 \times 10^{-6}$ mbar on top of the organic thin films. The device area was $\approx 1 mm^2$ for all devices. The general device structure used for our measurements was metal/organic semiconductor/metal (see Fig.~\ref{fig:DeviceSchematic}).

A schematic drawing of the device and the experiment are shown in Fig.~\ref{fig:DeviceSchematic}. The samples were mounted on the cold finger of a closed-cycle helium cryostat located between the poles of an electromagnet. The magnetoconductance ratio, $\Delta I/I$, was determined by measuring the current, $I$, at a constant applied voltage, $V$, for different magnetic fields, B. EL of the devices was measured with a photomultiplier tube that was shielded from the magnetic field using high-saturation mu-shield foil. All the reported data are for room temperature.

\section{Experimental Results, part I}

\subsection{\label{sec:level1} Magnetic Field Effect on Current and Electroluminescence}

\begin{figure}
 \includegraphics[width=\columnwidth]{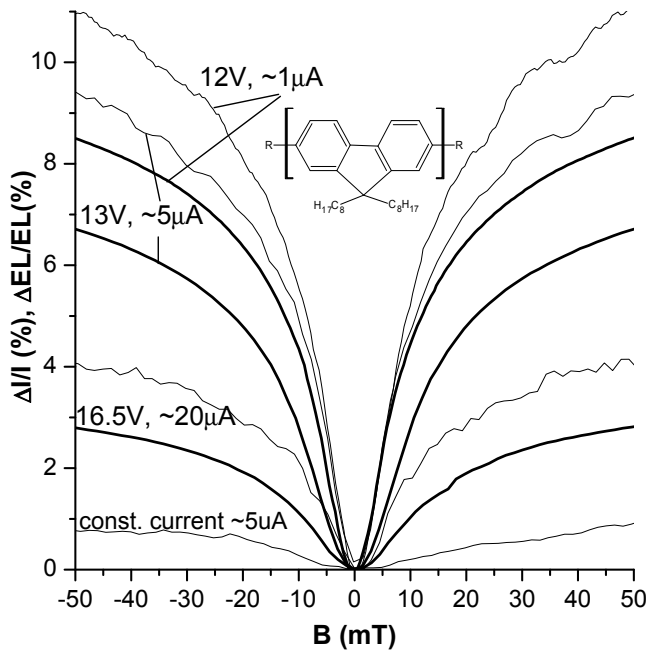}\\
 \caption{Magnetic field effect (MFE) on current (bold) and EL in a PEDOT/PFO($\approx$ 100 nm)/Ca device measured at several different constant voltages at room temperature. The MFE on EL measured at a constant current, 5 $\mu $A corresponding to $\approx$ 13 V, is also shown.}
 \label{fig:PFOMELI}
\end{figure}

\begin{figure}
\includegraphics[width=\columnwidth]{MELAlq3.prn}
\caption{MFE on current (bold) and EL in a PEDOT/Alq$_3$ ($\approx$100nm)/Ca device measured at several different constant voltages at room temperature. The MFE on EL measured at a constant current, 11 $\mu $A corresponding to $\approx 10 V$, is also shown.}
\label{fig:MELAlq3}
\end{figure}

OMAR devices have the unique property of showing large magnetoresistance while being also highly electroluminescent. Fig.~\ref{fig:PFOMELI} and Fig.~\ref{fig:MELAlq3} show $\Delta I/I$ and $\Delta EL/EL$ vs. B in a PEDOT/PFO/Ca and PEDOT/Alq$_3$/Ca device, respectively, measured at a constant voltage. These data show that the MFE exists both in the electric and luminescent measurements \emph{with comparable magnitude}. Note that the shape of $(\Delta I/I) (B)$ and $(\Delta EL/EL) (B)$ are equivalent and that both scale in the same manner upon changing V. Both effects therefore share a common origin. $EL$ and $I$ are related through:

\begin{equation}\label{equ:EL}
   EL \propto \eta I,
\end{equation}
where $\eta$ is the EL quantum efficiency. For the sake of the following discussion it is useful to break down the EL process into three steps \cite{Friend:1999}, and $\eta$ can accordingly be written as

\begin{equation}\label{equ:eta}
   \eta = \eta_{1} \eta_{2} \eta_{3},
\end{equation}
where $\eta_{1}$ is the fraction of the injected carriers that form electron-hole pairs, $\eta_{2}$ is the fraction of the total
number of excitons that are singlets, and $\eta_{3}$ is the singlet emission quantum efficiency.

There are two distinct possibilities regarding a potential explanation of the MFE on EL (MEL): (i) $I = I (B)$ and (ii) $\eta = \eta (B)$. In terms of the underlying mechanism, a type (i) model involves the charge carriers directly, whereas a type (ii) model is based on \emph{excitonic} mechanisms. Frankevich and coworkers \cite{Frankevich:1992} and Kalinowski and coworkers \cite{Kalinowski:June2003, Kalinowski:August2003} have previously used the excitonic pair mechanism model (i.e. a type (ii) model) for explaining their findings regarding the MEL effect. The question naturally arises whether their model can also explain OMAR. To answer this question we recast the pair-mechanism model in a form suitable to simultaneously examine the prediction of this model for $\Delta EL/EL$ and $\Delta I/I$. Before proceeding, however, we want to summarize the experimental evidence that supports the claim that the hyperfine interaction causes MEL and OMAR.

\section{Discussion, part I} \label{sec:discussionpartI}

\begin{figure}
 \includegraphics[width=\columnwidth]{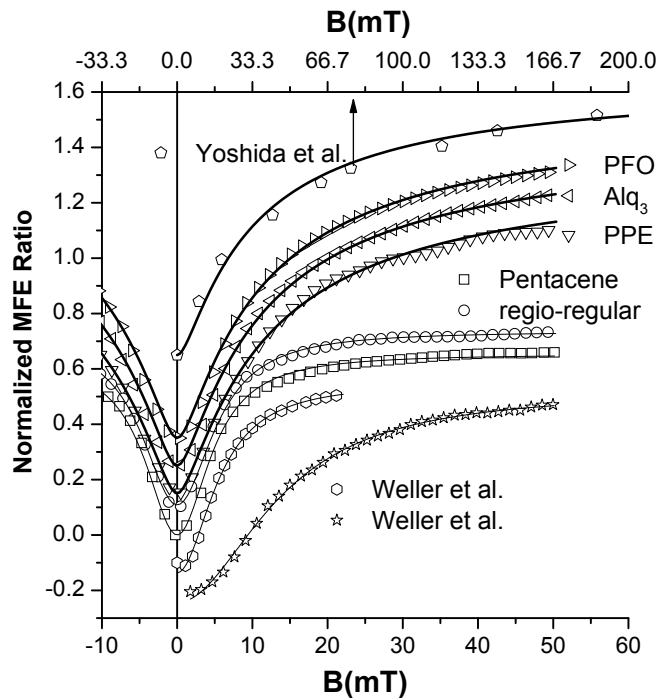}\\
 \caption{Normalized OMAR traces in PEDOT/Organic Layer/Ca devices, with PFO, Alq$_3$, poly(phenylene-ethynelene) (PPE), pentacene and regio-regular P3HT as the organic layers taken from Ref.\cite{Mermer:2005b}. For comparison, data on the MEL effect in a poly(phenylene-vinylene) OLED measured by Yoshida et al. \cite{Yoshida:2005} and that on the triplet photoyield in solutions of organic materials measured by Weller et al. \cite{Weller:1983,Staerk:1985} are also shown. The data are shown as scatter plots as detailed in the legend. The solid curves are fits using empirical laws of the forms $(B/(|B|+B_0))^2$ (thicker lines) and $B^2/(B^2+B_0^2)$(thinner lines). Please note that the data by Yoshida et al. and the corresponding fit refer to the upper x-axis, whereas all other data refer to the lower x-axis.}
 \label{fig:NormalizedDeltaRR}
\end{figure}

\begin{table}
 \centering
 \caption{Function and parameter values used for fits in Fig.~\ref{fig:NormalizedDeltaRR}}\label{tab:fittingresults}
 \begin{tabular}{|c|}
  \hline
  fitting function: $B^2/(B^2+B_0^2$) \\
  \hline
  \begin{tabular}{c|c}
    material name & $B_0/mT$ \\ \hline
    regio-regular P3HT & 5.1 \\
    pentacene & 5.8 \\
    Weller et al. 1 & 5.7 \\
    Weller et al. 2 & 13.7 \\
  \end{tabular} \\ \hline
  fitting function: $\left (B/(|B|+B_0) \right ) ^2$ \\ \hline
  \begin{tabular}{c|c}
    material name & $B_0/mT$ \\ \hline
    PFO & 5.4 \\
    Alq$_3$ & 5.4 \\
    PPE & 5.5 \\
    Yoshida et al. & 14 \\
  \end{tabular} \\
  \hline
 \end{tabular}
\end{table}

Fig.~\ref{fig:NormalizedDeltaRR} shows a summary of measured OMAR traces, taken from our previous publication \cite{Mermer:2005b}, in PEDOT/Organic Layer/Ca devices employing several different organic layers. The traces of the various data sets have been normalized to achieve a suitable graphical representation. The solid curves are fits using empirical laws of the forms $(B/(|B|+B_0))^2$ (thicker lines) and $B^2/(B^2+B_0^2)$ (thinner lines);  the values for the fitting parameter, $B_0$, are given in table~\ref{tab:fittingresults}. These empirical laws were introduced by us in a previous publication \cite{Mermer:2005b} and yield excellent one-parameter fits. For comparison, data on the MEL effect in a poly(phenylene-vinylenene) OLED measured by Yoshida et al. \cite{Yoshida:2005} and that on the triplet photoyield in organic solutions measured by Weller et al. \cite{Weller:1983,Staerk:1985} are also shown. Furthermore it is shown that the data by Yoshida et al. and Weller et al. can also be fitted accurately by our empirical laws suggesting a common origin of OMAR and other MFE data. Since Yoshida et al. and Weller et al. have interpreted their data in terms of the pair mechanism (hyperfine interaction) model, this suggests that OMAR is also caused by hyperfine interaction. As a matter of fact, Weller and coworkers \cite{Staerk:1985,Weller:1983} have directly shown through transient measurements of the delayed fluorescence caused by triplet-triplet annihilation in solutions of organic materials that a weak magnetic field can modulate the spin multiplicity within 10-20ns after optical excitation. Furthermore, Schulten et al. \cite{Schulten:1984} showed that the width of the MFE traces can be calculated from first principles based on the hyperfine coupling constants. However, Schulten et al. \cite{Schulten:1984} did not specify an analytical result for the dependence of the MFE on B and the origin of the \emph{simple, analytical} fitting formulas, $(B/(|B|+B_0))^2$ and $B^2/(B^2+B_0^2)$, is not yet well-established. We have therefore performed a simplified calculation of the dependence of the MFE on B (see Appendix) where we show that the origin of the empirical law, $B^2/(B^2+B_0^2)$, can be readily understood. We note that this formula is closely related to the Lorentzian function $B_0^2/(B^2+B_0^2)$. However, we do not yet understand the origin of the other empirical law, $B^2/(|B|+B_0)^2$.

\subsection{\label{subsec:width} What determines the width of OMAR traces?}

The simple model calculation presented in the Appendix gives the relation $B_0 = \left ( \sum_i a_{H,i}^2 \right )^{1/2}$ for the width of the OMAR traces, whereas Schulten's theory \cite{Schulten:1984} gives $B_0 = \sqrt{3} \left ( \sum_i a_{H,i}^2 \right )^{1/2}$ for nuclear spin 1/2. In either case, it should therefore be possible to calculate $B_0$ from published values for the hyperfine splitting, $a_H$ in the electron spin resonance spectra of the organic molecules (see e.g. Ref.~\cite{Lewis:1965}). As a matter of fact, the width of the Weller et al. \cite{Weller:1983} data numerically coincides with Schulten's formula with pretty good accuracy. However, for our OMAR data the experimental width is considerably greater than the calculated one. For pentacene, for example, we calculate $B_0=1mT$ and $B_0=1.8mT$ using our and Schulten's formula, respectively. The experimental value is however $B_0=5.8mT$. Indeed, the fact that the widths of the MFE traces in OLEDs are much wider than expected has already been recognized by Yoshida et al. \cite{Yoshida:2005}. Yoshida et al. therefore suggested that this is a result of lifetime broadening (see section~\ref{subsec:pairmechanism model}). Our experiments however exclude this possibility: The pair lifetime should be a sensitive function of temperature and should vary considerably between different materials in contradiction with experiment (see Fig.~\ref{fig:NormalizedDeltaRR} and Ref.~\cite{Mermer:2005b}). Although we have been unable to find a convincing explanation for this discrepancy, it is possible that hyperfine coupling with hydrogen nuclei of neighboring molecules in the densely packed films and anisotropic hyperfine coupling may account for (part of) the excess width. A further disagreement between theoretical expectation and experimental results is evident from the observation that $B_0$ is similar in both small molecules and polymers. This is unexpected because McConnell's relationship \cite{McConnell:1956} states that $a_{H,i}=Q\rho_i$, where $\rho_i$ is the spin density at nucleus, $i$, and $Q \approx 3 mT$ for conjugated molecules \cite{Lewis:1965}. For a polymer with $N$ repeat units we therefore have

\begin{eqnarray}
  a_H^2 &=& \sum (Q \rho_i)^2 = N \sum' (Q \rho_i)^2 \approx N \sum' (Q \rho'_i/N)^2 \\
  a_H & \propto & N^{-1/2},
\end{eqnarray}

where $\sum$ denotes a sum over all nuclei in the polymer, whereas $\sum'$ is a sum only over a single repeat unit. $\rho_i$ is the spin-density at nucleus, $i$, whereas $\rho'_i$ is the spin density at nucleus, $i$, in the corresponding monomer. This result implies that the MFE cones should be considerably narrower in polymers than in small molecules such as Alq$_3$, in contradiction with the experimental results (see Fig.~\ref{fig:NormalizedDeltaRR}).

\subsection{\label{subsec:implications} Potential implication of hyperfine coupling to organic spintronics application}

In the preceding paragraphs we have summarized some of the experimental evidence for the importance of hyperfine coupling in organics. Furthermore, the discovery of OMAR and its possible relation to hyperfine interaction illustrates the importance of the study of spin-dynamics in relation to transport phenomena. In particular, we believe that the existence of hyperfine interaction has fundamental implications to the currently emerging field of organic spintronics \cite{Dediu:2002,Xiong:2004}. Since hyperfine coupling leads to time-evolution of the electron spin and since the local nuclear spin configuration is different for each molecule in the film, hyperfine coupling will lead to spin-decoherence. The relevant decoherence time-scale, $T_2$ is given by

\begin{equation}\label{equ:decoherence}
    T_2 \approx \frac{g \mu_B}{\hbar} B_0 \approx 1ns
\end{equation}

Assuming the drift term to be dominant in OLEDs, we therefore obtain for the spin-transport length, $\lambda$

\begin{equation}\label{equ:spintransport}
    \lambda \approx \mu F T_2 \approx 1 \AA
\end{equation}

where we have used $\mu = 10^{-4}cm^2(Vs)^{-1}$ for a typical mobility value and $F=10^5 Vcm^{-1}$ for a typical value for the electrical field. This result clearly shows that hyperfine interaction may seriously limit the spin-transport properties of organic semiconductors. Of course, the effect of hyperfine coupling can be switched off by applying $B > B_0$, a property which may have some use in spintronics applications.

\subsection{\label{subsec:pairmechanism model} Excitonic pair mechanism model }

\begin{figure}
\includegraphics[width=\columnwidth]{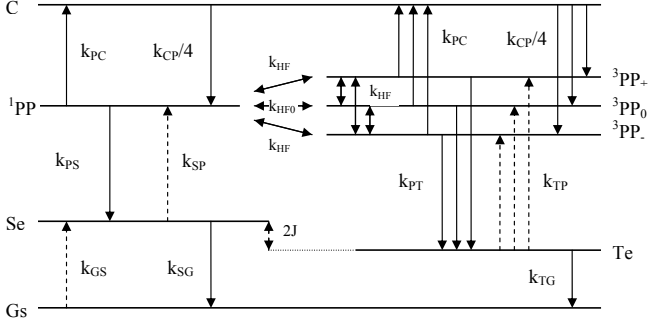}
\caption{\label{fig:MFEmodel}The schematic energy level diagram illustrates the simplest possible pair mechanism model which includes three different species:(i) free charges with population $C$, (ii) polaron pairs PPs, and (iii) singlet $Se$ and triplet excitons $Te$ with large exchange interaction, J. The various transition rates are indicated.}
\end{figure}

Next, we examine whether OMAR can be explained by the \emph{excitonic} pair mechanism model that was applied to MFE in OLEDs by Frankevich \cite{Frankevich:1992} and Kalinowski \cite{Kalinowski:August2003}. When electrons and holes are injected from the cathode and anode into the organic layer, they form negative and positive polarons, respectively \cite{Heeger:1988,Vardeny:1997}. As long as the distance between positive and negative carriers is larger than the Coulomb capture radius, $r_{C}$, they do not feel each other's attraction, and we refer to them as free charges. In order for the pair mechanism model \cite{Frankevich:1992,Kalinowski:August2003} to work we assume that at separations less than $r_{C}$, the carriers are organized in bound pairs, which we refer to as a polaron pair (PP). At separations less than the single particle wavefunction extend, the exchange interaction becomes important and the pair has to be represented by a single, properly symmetrized, wavefunction. We refer to this as an exciton. We note that the pair mechanism model is an example of a spin-dependent effect that does not require (thermal) spin-polarization. One way of understanding this is to realize that, since the carriers form pairs, the "ensemble" consists of two spins only and very large "effective spin-polarization" is automatically achieved. The formation of pairs is therefore essential to this model.

Closely following the treatment by Frankevich \cite{Frankevich:1992} we now formulate rate equations that govern the population of free charges, polaron pairs and singlet and triplet excitons. The relevant levels and transition rates are shown in Fig.~\ref{fig:MFEmodel}. $^{1}PP$, $^{3}PP_{0}$, $^{3}PP_{+}$, $^{3}PP_{-}$ denote the pair populations, where the superscript denotes the multiplicity, whereas the subscript denotes the spin-projection. Because of negligible exchange interaction in $PP$, the four $PP$ states are degenerate for zero applied B-field. The basic idea of the pair mechanism model is that the PPs may undergo a spin motion induced by the hyperfine interaction \cite{Schulten:1978,Knapp:1979,Schulten:1984,Steiner:1989} (see Appendix). We denote the rate of conversion between $^1PP$ and $^3PP_0$ as $k_{HF_0}$, and that between other PP as $k_{HF}$. If B=0, then $k_{HF}=k_{HF_0}$. An applied B-field leads to Zeeman splitting, $\Delta E$ between levels of different spin-projection. If $\triangle E>\hbar k_{HF_0}$, then $k_{HF}=0$, however, $k_{HF_{0}}$ remains unchanged. In summary of this paragraph, the simplest possible model for explaining the magnetic field effect of I and EL requires rate equations for three different species, namely (i) free charges that carry the current, (ii) pairs, such that it becomes meaningful to talk about singlet and triplet states, and (iii) excitons that, by virtue of their large exchange energy, furnish the model with spin-dependent pair recombination rates. We note that pairs and excitons do not contribute to the current, because of their overall neutrality. The various rate equations are given by:

\begin{eqnarray}
&&G_{C}+k_{PC}\left(\sum (^{i}PP)\right) -k_{CP}C = 0\label{eqn2}\\
&&\frac{1}{4}k_{CP}C+k_{HF_{0}}(^{3}PP_{0}-^{1}PP)+k_{HF}(^{3}PP_{+}-^{1}PP)\nonumber\\
&&+k_{HF}(^{3}PP_{-}-^{1}PP)-(k_{PC}+k_{PS})^{1}PP=0\label{eqn3}\\
&&\frac{1}{4}k_{CP}C+k_{HF_{0}}(^{1}PP-^{3}PP_{0})+k_{HF}(^{3}PP_{+}-^{3}PP_{0})\nonumber\\
&&+k_{HF}(^{3}PP_{-}-^{3}PP_{0})-(k_{PC}+k_{PT})^{3}PP_{0}=0\label{eqn4}\\
&&\frac{1}{4}k_{CP}C+k_{HF}(^{1}PP-^{3}PP_{\pm})+k_{HF}(^{3}PP_{0}-^{3}PP_{\pm})\nonumber\\
&&+k_{HF}(^{3}PP_{\mp}-^{3}PP_{\pm})-(k_{PC}+k_{PT})^{3}PP_{\pm}=0\label{eqn5}
\end{eqnarray}

The definitions of the various transition rates are given in Fig.~\ref{fig:MFEmodel}. The first equation is the rate equation for the free charges, C. $G_C$ is the generation rate for C, which is equal to the rate of carrier injection minus the rate of emission of carriers at the electrodes. The next three equations are the spin-dependent rate equations for PPs. We neglected the upwards transitions $k_{SP}, k_{TP}, k_{GS}$ (dashed arrows in Fig.~\ref{fig:MFEmodel}) to simplify the rate equations. We assume that $G_{C}$ is constant as long as the voltage remains fixed, in particular it is independent of B \cite{Mermer:2005b}. We obtain the following solutions to the rate equations:

\begin{eqnarray}
\left.\frac{\triangle
I}{I}\right|_{V} &=&\eta_{1}\frac{\frac{k_{PC}}{k_{PT}}(1-r)^2}{(4\frac{k_{PC}}{k_{PT}}+r+3)[\frac{k_{PC}}{k_{PT}}(r+3)+2(r+1)]}\label{eqn7}\\
\left.\frac{\triangle EL}{EL}\right|_{V}&=&\frac{1-r}{\frac{k_{PC}}{k_{PT}}(r+3)+2(r+1)}\label{eqn8}\\
R&=&\frac{\left.\frac{\triangle I}{I}\right|_{V}}{\left.\frac{\triangle EL}{EL}\right|_{V}} = \eta_{1}\frac{\frac{k_{PC}}{k_{PT}}(1-r)}{4\frac{k_{PC}}{k_{PT}}+r+3}\label{eqn9}\\
\left.\frac{\triangle
EL}{EL}\right|_{I}&=&\frac{1-r}{2[2\frac{k_{PC}}{k_{PT}}+r+1]}\label{eqn10}
\end{eqnarray}

where we have used $EL \propto$ $^{1}PP$ and $I \propto C$; $r \equiv k_{PS}/k_{PT}$ \cite{Wohlgenannt:2001}. The notations $|_V$ and $|_I$ refer to measurements where the voltage, V, or the current, I, were kept constant, respectively. We have used $k_{HF_0} \gg k_{PS}, k_{PT}, k_{PC}$ (see section~\ref{subsec:width}).

We will now show that these results are in clear contradiction with experiments. Equ.~(\ref{eqn7}) shows that $\triangle I/I|_{V}$ is always positive in this model in contradiction with the experimental results where both positive and negative magnetoconductance is observed \cite{Mermer:2005b}. Furthermore, it is seen that $\triangle I/I|_{V}$ is a second order effect, whereas $\triangle EL/EL|_{V}$ appears in first order in $(1-r)$. This contradicts the experimental observation (see Figs.~\ref{fig:PFOMELI} and \ref{fig:MELAlq3}) that there exists a fixed ratio, $R$, between the two effects. The model result for R (Equ.~(\ref{eqn9})), however, is a function of $r$ and $k_{PC}/k_{PT}$. In the present case of polyfluorene and Alq$_3$ OLEDs, the positive sign of R can be brought into agreement with the model only if $r<1$. So in the following discussion we will limit ourselves to this case. For the sake of discussion, we will study two limits: $k_{PC}/k_{PT}\gg 1$ and $k_{PC}/k_{PT}\ll 1$. If $k_{PC}/k_{PT}\gg 1$, (\ref{eqn7}) (\ref{eqn8}) (\ref{eqn10}) can be
simplified as:

\begin{eqnarray}
\left.\frac{\Delta
I}{I}\right|_{V}=\eta_{1}\frac{(1-r)^2}{4\frac{k_{PC}}{k_{PT}}(r+3)};
\left.\frac{\Delta EL}{EL}\right
|_{V}=\frac{1-r}{\frac{k_{PC}}{k_{PT}}(r+3)}\nonumber\\
\left.\frac{\Delta EL}{EL}\right
|_{I}=\frac{1-r}{4\frac{k_{PC}}{k_{PT}}}\label{eqn11}
\end{eqnarray}

If $k_{PC}/k_{PT}\ll 1$, (\ref{eqn7}) (\ref{eqn8}) (\ref{eqn10}) can
be simplified as:

\begin{eqnarray}
\left.\frac{\Delta I}{I}\right
|_{V}=\eta_{1}\frac{\frac{k_{PC}}{k_{PT}}(1-r)^2}{2(r+1)(r+3)};
\left.\frac{\Delta EL}{EL}\right|_{V}=\frac{1-r}{2(r+1)}\nonumber \label{equ:ELV}\\
\left.\frac{\Delta EL}{EL}\right|_{I}=\frac{1-r}{2(r+1)}\label{equ:ELI}
\end{eqnarray}

From these results it follows that $\left. \frac{\Delta I}{I}\right|_{V}$ can be large only if $k_{PC} \approx k_{PT}$. Whereas this condition could be accidentally satisfied in a single material at a single temperature, it cannot hold true at all temperatures in different materials. The model's conclusion that $\left. \frac{\Delta I}{I}\right|_{V}$ is always small results because the free charges, the only current carrying species in the model, do not directly participate in the spin-dependent reactions or the hyperfine transitions. In particular, when $k_{PC}/k_{PT}\gg 1$ most polarons exist as free charges and the pair concentration and therefore the MFE are small. When $k_{PC}/k_{PT}\ll 1$ the pair concentration and therefore the MFE are large, but the large change in the pair population does not affect the current because the pairs do not dissociate. Equs.~(\ref{eqn11}) and (\ref{equ:ELV}) highlight another contradiction between model and experiment: In the present model it makes little difference whether $\Delta EL/EL$ is measured while the voltage or current are kept constant, whereas the experimental results (Figs.~\ref{fig:PFOMELI} and \ref{fig:MELAlq3}) showed that $\Delta EL/EL|_{V}$ is about ten times bigger than $\Delta  EL/EL|_{I}$.

\section{Experimental Results, part II}

\begin{figure}
 \includegraphics[width=\columnwidth]{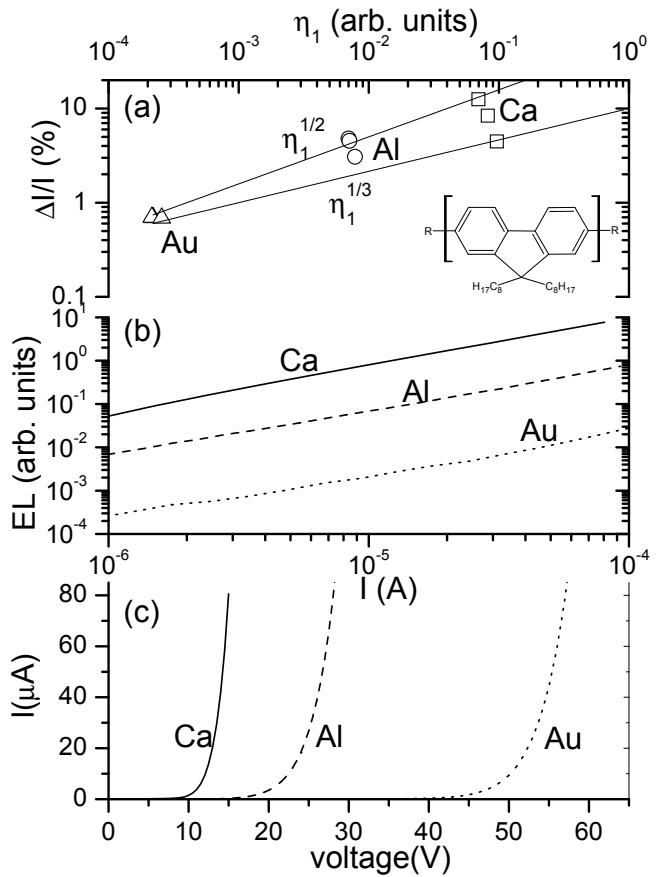}\\
 \caption{a) Magnitude of the magnetoconductance, $\Delta I/I$, in PFO (see inset) as a function of the exicton/carrier ratio $\eta_1$ at $B=100 mT$. b) EL as a function of current. c) Current-voltage (I-V) characteristics. All data were obtained at room temperature.}
 \label{fig:PFOCathodes}
\end{figure}

\begin{figure}
 \includegraphics[width=\columnwidth]{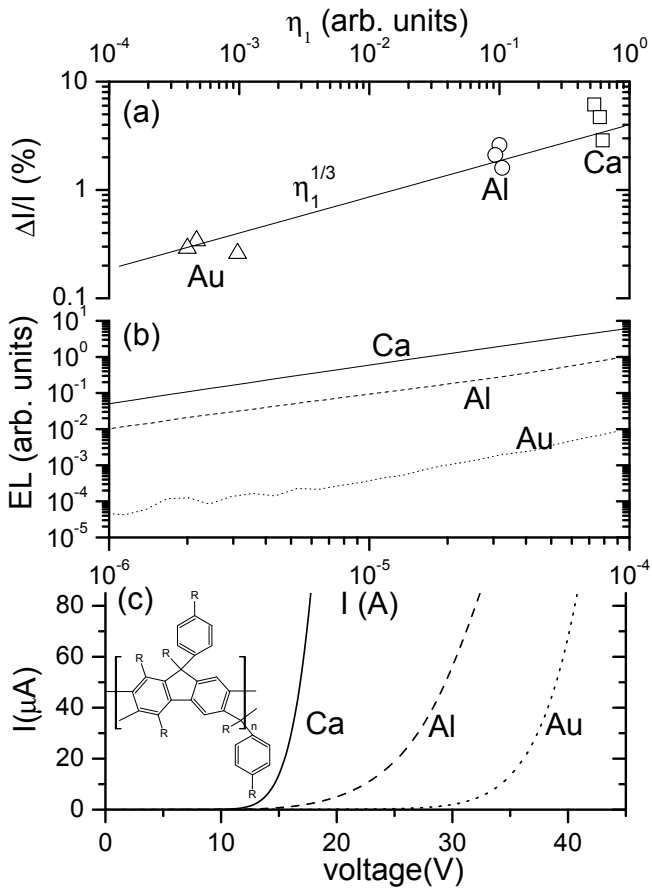}\\
 \caption{a) Magnitude of the magnetoconductance, $\Delta I/I$, in MeLPPP (see inset) as a function of the exicton/carrier ratio $\eta_1$ at $B=100 mT$. b) EL as a function of current. c) Current-voltage (I-V) characteristics. All data were obtained at room temperature.}
 \label{fig:MeLPPPCathodes}
\end{figure}

In this section we determine experimentally whether OMAR is related to an excitonic effect or not. This is possible, since $\eta_1$ that appears in Equ.~(\ref{eqn8}) can be varied experimentally by several orders of magnitude. By varying the corresponding electrode materials one can control the injection of minority charge carriers, so that the exciton formation can be tuned with ideally minimal effect on the current density, which is determined mostly by the majority carriers. This can be easily achieved in hole-dominated PFO devices by choosing cathode (top electrode) materials with different work functions, but it is difficult in practice to fabricate electron-only Alq$_3$ devices with different anode (bottom electrode) materials, because efficient electron injection requires reactive metals such as Ca. We found, e.g. in a Ca/Alq$_3$/Ca device, that the bottom Ca electrode oxidizes quickly before and/or after evaporating Alq$_3$ on top of it. So, instead of Alq$_3$ devices we studied a second polymer, MeLPPP (see Fig.~\ref{fig:MeLPPPCathodes}, inset) in addition to PFO, to show that the conclusions we draw are not limited to a particular choice of polymer. MeLPPP is a suitable choice, because it shows, like PFO devices, large OMAR as well as intense EL.

Three types each of PFO and MeLPPP devices were fabricated and measured using Ca, Al and Au as cathode materials. We note that whereas only one set of data points for each device is shown, the reported experiments were repeated several times and very reproducible results were obtained. Figs.~\ref{fig:PFOCathodes} and \ref{fig:MeLPPPCathodes} show the current-voltage (I-V) characteristics (Figs.~\ref{fig:PFOCathodes}c and \ref{fig:MeLPPPCathodes}c), the measured EL intensity as a function of I (Figs.~\ref{fig:PFOCathodes}b and \ref{fig:MeLPPPCathodes}b), and the magnitude of $\Delta I/I$ as a function of $\eta_{1}$ (Figs.~\ref{fig:PFOCathodes}a and \ref{fig:MeLPPPCathodes}a). The exciton/carrier ratio, $\eta_1$ was inferred from the data shown in panels b) where we found that the magnitude of EL, at a given current, in Ca devices is about one order of magnitude larger than that in Al devices, and nearly three orders of magnitude larger than in Au devices. This is evidence of the increasing difficulty of electron injection from the cathode into the polymer going from low work-function Ca to high work-function Au. This is well-known \cite{Friend:1999} to result from the mismatch of the cathode work function and the polymer's conduction band. Correspondingly, $\eta$ is one (three) orders of magnitude lower in Al (Au) devices compared to Ca devices. Since, $\eta_2$ and $\eta_3$ in Equ.~(\ref{equ:eta}) are properties of the excitons \cite{Baldo:1999} in each organic material, and they do therefore not depend on the carrier injection efficiency (hence the cathode materials), we simply use $\eta$ as a measure of the exciton/carrier ratio, $\eta_1$, of the various devices.

Most importantly, Figs.~\ref{fig:PFOCathodes}a and ~\ref{fig:MeLPPPCathodes}a show that the magnitude of $\Delta I/I$ increases as $\eta_{1}$ increases in both PFO and MeLPPP devices (results for three choices of the current are shown for each device). At first sight this trend seems to confirm an excitonic origin of OMAR. However, closer inspection shows that the $\eta_{1}$ dependence is much weaker than expected. Whereas a linear dependence on $\eta_1$ is expected for an excitonic effect, we find $\Delta I/I \propto \eta_{1}^{\alpha}$, with $\alpha$ ranging from 1/3 to 1/2. If, however, OMAR is not related to an excitonic effect, but caused by a direct effect of B on the (mobility of the) free carriers, we would have expected no dependence at all, which does not match the measurement either. However, the observed weak dependence can easily be attributed to several possible factors: It is possible that the interface resistance of polymer/Au is larger than that of polymer/Ca resulting in additional resistance that is not subject to OMAR. Previous studies \cite{Campbell:2001} report that Au indeed leads to a non-Ohmic top contact, possibly because the wetting of Au and therefore the physical contact is inferior, or because Au deposition, which has to be evaporated at a much higher temperature than Ca, leads to damage of the underlying polymer surface. Moreover, in unipolar devices space-charge limited current conditions occur which are possibly unfavorable for OMAR. In bipolar devices, however, the space charge of the two carrier types partially cancel each other. The fact that changing the cathode material leads to additional effects, rather than merely changing $\eta_1$, is shown in Figs.~\ref{fig:PFOCathodes}c and \ref{fig:MeLPPPCathodes}c where it is seen that Au cathode devices show a significantly increased device resistance, probably due to the above mentioned parasitic resistance contributions. We have however been unable to come up with any probable or improbable explanation how the dependence on $\eta_1$ could be sub-linear in an excitonic model. We could imagine superlinear behavior, if the spin-dependent reactions are bimolecular. Our measurements therefore show that OMAR is most likely not related to an excitonic effect.

\section{Discussion, part II}

The excitonic pair mechanism model fails to explain the comprehensive experimental data, but it still succeeds to account for several experimental observations (namely the magnitude and "universality" of $B_0$) that would otherwise be hard to explain with any model that does not involve the nuclear spin. So, it appears that a model is demanded, which retains some of the features of this model, namely the hyperfine interaction coupled to spin-dependent interactions, but applies them to a spin-dependent reaction that directly influences the mobility. The suggestion by Frankevich \cite{Frankevich:1992} to consider spin-dependent bipolaron formation \cite{Heeger:1988} appears promising. In the following we will demonstrate that the \emph{assumption} that B acts directly on the mobility, $\mu$, correctly reproduces the experimental data. We are however presently not able to identify or formulate a specific mechanism with the required properties. Such a theory lies far beyond the scope of the present paper and we hope that the scientific community will attempt to formulate such a theory.


We now wish to discuss the dependence of EL and I on the mobility. The most general formula for for the current density, $J$, is:

\begin{equation}\label{equ:current}
   J=e(p\mu_p - n\mu_n)F
\end{equation}

where $e$ is the elementary charge, $p$ and $n$ are the density of positive and negative polarons, respectively and $\mu_p$ and $\mu_n$ are their respective mobilities, $F$ is the electric field. All these quantities may in general depend on the position, $x$. In the case of unipolar transport, $J$ in OLEDs is often assumed to be given by in the Space Charge Limited Current (SCLC) formula:

\begin{equation}\label{equ:SCLC}
   J=\frac{9}{8}\epsilon \epsilon_{0}\mu\frac{V^2}{d^3},
\end{equation}

where $V$ is the applied voltage and $d$ is the device thickness, $\epsilon$ and $\epsilon_0$ have their usual meaning. This formula holds true only in the trap-free limit, in the case of traps the term $\left ( \frac{V}{d} \right )$ in Equ. (\ref{equ:SCLC}) must be replaced by a more general function, $f \left ( \frac{V}{d} \right )$ \cite{Pope:1999}. In the following we will assume that the mobility of one type of carrier greatly exceeds that of the other, specifically $\mu_p \gg \mu_n$. This assumption is usually fulfilled in OLEDs.

The relation for EL depends on the the relative magnitude of recombination time ($\tau_{rec}$) and transit time ($\tau_{t}$) \cite{Kalinowski:July1999}. The ratio $\tau_{rec}/(\tau_{rec}+\tau_{t})$ is a measure of the probability that the carrier will transit the device without recombining with another carrier. For injection-controlled EL ($\tau_{t}<\tau_{rec}$) we obtain:
\begin{eqnarray}
EL & \propto & \gamma \frac{j_{n}j_{p}}{\mu_{n} \mu_{p}} \label{IC1}\\
j_{total} & \approx & j_{p}\label{IC2} \\
\frac{\Delta EL}{EL} & = & \frac{\Delta I}{I} = \frac{\Delta \mu_p}{\mu_p} \label{equ:IC4}
\end{eqnarray}

$\gamma$ is the electron-hole recombination constant, which is usually assumed to be given by the Langevin recombination formula \cite{Pope:1999}, $\gamma=e(\mu_{n}+\mu_{p})/\epsilon \epsilon_{0}$. Equ.~(\ref{equ:IC4}) shows that the experimentally observed intimate relation between $\frac{\Delta EL}{EL}$ and $\frac{\Delta I}{I}$ naturally follows in a type (i) model. However, result Equ.~(\ref{equ:IC4}) implies that the two effects should be strictly of equal magnitude, which is not quite the case experimentally (see Figs.~\ref{fig:PFOMELI} and \ref{fig:MELAlq3}) where $\frac{\Delta EL}{EL}$ is up to twice as big as $\frac{\Delta I}{I}$. A more general relationship is therefore required. For the case $\tau_{t}^{n}>\tau_{rec}>\tau_{t}^{p}$ we obtain:
\begin{eqnarray}
EL & \propto & j_{n} \label{VC1} \\
j_{total} & \approx & j_{p}\label{VC2} \\
\frac{\Delta EL}{EL} & = & \frac{\Delta \mu_n}{\mu_n}; \frac{\Delta I}{I} = \frac{\Delta \mu_p}{\mu_p}\label {VC3}
\end{eqnarray}

Furthermore, for volume controlled EL ($\tau_{t}>\tau_{rec}$ \cite{Kalinowski:July1999}), which requires $\mu_p \approx \mu_n$, we obtain:
\begin{eqnarray}
\frac{\Delta EL}{EL} &=& \frac{\frac{\Delta \mu_n}{\Delta \mu_p}+\left ( \frac{\mu_n}{\mu_p} \right )^2}{\frac{\mu_n}{\mu_p} (\frac{\mu_n}{\mu_p}+1)} \frac{\Delta \mu_p}{\mu_p}\label {VC9}\\
 \frac{\Delta I}{I} &=& \frac{\frac{\Delta \mu_n}{\Delta \mu_p}+\frac{\mu_n}{\mu_p}}{2\frac{\mu_n}{\mu_p}} \frac{\Delta
 \mu_p}{\mu_p}\label {VC10}
\end{eqnarray}

In summary, a pair mechanism model that acts directly on the current carrying species naturally implies the observed intimate relationship between $\Delta EL/EL$ and $\Delta I/I$. Furthermore it implies the experimental observation that $\Delta EL/EL|_I \ll \Delta EL/EL|_V$ (see Figs.~\ref{fig:PFOMELI} and \ref{fig:MELAlq3}).

\section{Conclusion}

We have explored the possibility that hyperfine interaction causes the recently discovered organic magnetoresistance effect using both experimental data and theoretical models. Other groups have previously explained magnetic field effects on excitonic processes in organics using the hyperfine pair mechanism model. We show that both kinds of data can be fitted using the same empirical laws, either of the form $B^2/(B^2+B_0^2)$ or $B^2/(|B|+B_0)^2$ dependent on material. The only fitting parameter, $B_0$, assumes values that, at least at first sight, seem typical of hyperfine interaction. This suggests that OMAR is caused by hyperfine interaction as well. We succeeded to "derive" the empirical law $B^2/(B^2+B_0^2)$ from the standard hyperfine Hamiltonian. We also showed that hyperfine interaction may seriously limit the spin transport length, which is of primary importance in spintronics applications.

In order to further test the hyperfine interaction hypothesis, we examined a pair mechanism model, suggested previously by other authors which considers the electron spin precession caused by the hydrogen nuclear magnetic field. However, we found several fundamental contradictions with the existing experimental data. This model yields only a small and necessarily positive magnetoconductance effect, whereas experimentally a large effect is found, either of positive or negative sign. It is found that whereas $\Delta EL/EL$ is a first order effect, $\Delta I/I$ appears only in second order, in contradiction to the experimental observation that they are of similar magnitude. In addition, this model cannot account for the experimental observation that $\Delta EL/EL$ measured at constant voltage greatly exceeds the same effect when measured at constant current. In summary, we find that the excitonic pair mechanism model failed to explain OMAR, even only qualitatively. We trace the origin of the failure of the model to the fact that it considers the spin-dynamics of polaron pairs, which, in first order, do not contribute to the current.

By varying the injection efficiency for minority carriers in the devices, we show experimentally that $\Delta I/I$ is only weakly dependent on the EL quantum efficiency: $\Delta I/I \propto \eta_1 ^{\alpha}$, with $\alpha$ ranging from 1/3 to 1/2. This dependence is unexpectedly weak if the effect were of an excitonic origin. This observation strengthens the conclusion from the modelling that OMAR is not due to an excitonic effect. 

Making the (presently unjustified) assumption that B acts directly on the carrier mobility, we found that the qualitative features of both OMAR and the magnetic field effect on the electroluminescence naturally follow from this assumption. Therefore, we believe that a pair mechanism model acting on the current carriers directly is a promising direction for a future theory of OMAR. Spin-dependent bipolaron formation is a possible scenario.

\section{Appendix: Lorentzian shape of magnetic field effect traces deduced from the hyperfine Hamiltonian}

Here we will show that the empirical fitting formula $\Delta I/I \propto B^2/(B^2+B_0^2)$ can be substantiated through a simple calculation to be presented in the following. The model considers the standard hyperfine Hamiltonian,

\begin{equation}\label{equ:Hamiltonian}
  \hat{H} = \omega_0 \hat{S}_{z}+ \frac{a}{\hbar} \hat{\overrightarrow{S}}\hat{\overrightarrow{I}},
\end{equation}

containing the electronic Zeeman energy and the hyperfine interaction between a single electronic and nuclear dipole. $\omega_0=\frac{g\mu_BB}{\hbar}$, where $g \approx 2$ is the electronic g-factor and $\mu_B$ is the electronic Bohr magneton, $\overrightarrow{S}$ and $\overrightarrow{I}$ are the electronic and nuclear spin (assumed to be 1/2), respectively, and $a$ is a measure (in units of frequency) of the hyperfine interaction strength. The z-axis is chosen to coincide with the direction of B. The Hamiltonian will be written in matrix form where we use the following basis vectors: $|\Uparrow\uparrow>$, $|\Downarrow\uparrow>$, $|\Uparrow\downarrow>$ and $|\Downarrow\downarrow>$. The boldface arrow denotes the z-component of the electronic spin, whereas the second arrow denotes that of the nuclear spin. We obtain the following result:

\begin{equation}\label{equ:Hmatrix}
\hat{H}=\hbar\left[%
\begin{array}{cccc}
 \frac{\omega_0}{2}+\frac{a}{4} & 0 & 0 & 0 \\
 0 & -\frac{\omega_0}{2}-\frac{a}{4} & \frac{a}{2} & 0 \\
 0 & \frac{a}{2} & \frac{\omega_0}{2}-\frac{a}{4} & 0 \\
 0 & 0 & 0 & -\frac{\omega_0}{2}+\frac{a}{4} \\
\end{array}%
\right]
\end{equation}

It is evident from the form of $\hat{H}$ that $|\Uparrow\uparrow>$ and $|\Downarrow\downarrow>$ are eigenstates and therefore do not evolve with time other than through the trivial phase factor. However, $|\Downarrow\uparrow>$ and $|\Uparrow\downarrow>$ are mixed with each other through the off-diagonal matrix element. For simplicity we will now consider the time evolution of the $|\Uparrow\downarrow>$ state only. It turns out that a calculation of the time evolution of the most general state vector and subsequent averaging over all electronic and nuclear spin orientations leads to similar results. To obtain the time evolution operator we perform a matrix exponentiation:

\begin{equation}\label{equ:exponentiation}
  \hat{U} = exp \left (\frac{\hat{H}t}{i\hbar} \right)
\end{equation}

Next we calculate the expectation value of $\hat{S}_{z}$ as a function of time:

\begin{equation}\label{equ:expectationvalue}
  S_z(t) = <\Uparrow\downarrow \hat{U}|\hat{S}_z|\hat{U} \Uparrow\downarrow>
\end{equation}

We obtain the following result (in units of $\hbar/2$):

\begin{equation}\label{equ:result}
  S_z(t) = \frac{\omega_0^2}{\omega_0^2+a^2} + \frac{a^2}{\omega_0^2+a^2} cos \sqrt{\omega_0^2+a^2} t
\end{equation}

Next we consider the case of a pair of spins each of which are subject to a separate Hamiltonian of form Equ.~(\ref{equ:Hamiltonian}). For simplicity we will treat the time-evolution of the initial state $|\Uparrow\uparrow>_1|\Uparrow\downarrow>_2$ only. Since the first spin is in an eigenstate it will not evolve with time, whereas the second spin's time evolution will be governed by Equ.~(\ref{equ:exponentiation}). We obtain therefore for the total spin (in units of $\hbar$):

\begin{eqnarray}\label{equ:result}
  S^{1+2}_z(t) & = & \frac{1}{2}\left (1+\frac{\omega_0^2}{\omega_0^2+a^2} + \frac{a^2}{\omega_0^2+a^2} cos \sqrt{\omega_0^2+a^2} t \right ) \\
  & = & \frac{1}{2}+\frac{1}{2}\left ( p_{P}-p_{AP} \right ),
\end{eqnarray}

where $p_{P}$ and $p_{AP}$ are the probability for finding the pair in a parallel or antiparallel state, respectively. It is seen that $S^{1+2}_z(t)$ oscillates with time and that the peak-to-peak modulation depth is given by $\frac{a^2}{\omega_0^2+a^2}$. At large B, $S^{1+2}_z(t)$ remains close to $1$ at all times although it performs a high-frequency (but small-amplitude) oscillation. At small B, the frequency of the oscillation becomes smaller but its amplitude increases. The question arises whether the frequency or the amplitude of the oscillation is the correct measure for the "spin-flip efficiency". Because experiment shows (see section~\ref{subsec:width}) that the oscillation frequency is much larger than the pair recombination rate, $\gamma$, it is the time average of $S^{1+2}_z(t)$ that enters into the transition rate, specifically we may write:

\begin{eqnarray}
 \gamma &=& \overline{p_{P}} \gamma_{P} + \overline{p_{AP}} \gamma_{AP} \\
 &=& \left ( \frac{1}{2}+\frac{\omega_0^2}{2(\omega_0^2+a^2)} \right ) \gamma_{P} + \frac{a^2}{2(\omega_0^2+a^2)} \gamma_{AP} \\
  \frac{\Delta \gamma}{\gamma} & \equiv & \frac{\gamma (B)-\gamma (B=0)}{\gamma (B=0)} = \frac{\omega_0^2}{\omega_0^2+a^2} \frac{\gamma_P-\gamma_{AP}}{\gamma_P+\gamma_{AP}} \label{equ:final}
\end{eqnarray}

$\gamma_P$ and $\gamma_{AP}$ are the recombination rates for parallel and antiparallel pairs, respectively. For a state initially in an antiparallel state $\gamma_{P}$ and $\gamma_{AP}$ have to be exchanged in Equ.~(\ref{equ:final}).

Finally we relate our results to the experimentally reported values, $a_H=\frac{\hbar a}{g\mu_B}$, for the hyperfine coupling strength:

\begin{equation}\label{equ:appendixfinal}
  MFE \propto \frac{B^2}{B^2+a_H^2}
\end{equation}

Schulten and coworkers \cite{Schulten:1978} have shown that if the electron spin interacts with a large number of nuclear spins (as in the case of organic semiconductors), then the individual $a_{H,i}$ have to be added in a random-walk-like manner. The final relation for $B_0$ is therefore:

\begin{equation}\label{equ:B0}
  B_0 = \left ( \sum_i a_{H,i}^2 \right )^{1/2}
\end{equation}

\bibliography{Bibliography}

\end{document}